\documentclass[10pt,showpacs,preprintnumbers,amsmath,amssymb]{revtex4}
\usepackage{mathptmx}
\usepackage{units}
\usepackage{bbold}
\usepackage{graphicx}
\usepackage{dcolumn}
\usepackage{bm}
\usepackage{bbm}
\usepackage{slashed}

\begin{document}

\title{Four-dimensional conformal field theory using quaternions}
\vspace{2cm}
\author{Sergio Giardino}
\email{p12@ubi.pt}
\affiliation{ Departamento de F\'isica \& Centro de Matem\'atica e Aplica\c c\~oes, Universidade da Beira Interior\\
Rua Marqu\^es D'\'Avila e Bolama 6200-001 Covilh\~a, Portugal}

\begin{abstract}
\noindent We build a four-dimensional quaternion-parametrized conformal field theory (QCFT) using quaternion holomorphic
functions as the generators of quaternionic conformal transformations. Taking  
the two-dimensional complex-parametrized conformal field theory (CCFT) as our model, 
we study the stress tensor, the conserved charge, the symmetry 
generators, the quantization conditions and several operator product expansions (OPE's). Future applications are also addressed.
\end{abstract}

\maketitle

\section{Introduction}

A conformal field theory is a framework used in many applications in physics. These models 
may be built for any dimension, but the two-dimensional conformal field theory \cite{Ginsparg:1988ui,Blumenhagen:2009zz} is the most widely used,
 and it involves many 
features that explain this widespread use. The infinite-dimensional conformal algebra and its central extension
are certainly important applicability factors. Another feature is the parametrization of 
two-dimensional CFT using complex numbers, and thus the employment of the full machinery of complex analysis in order to build a 
complex-parametrized conformal field theory (CCFT).

On the other hand, four-dimensional conformal field theories have attracted considerable attention for recent applications in high-spin
models \cite{Rattazzi:2008pe,Vos:2014pqa,Elkhidir:2014woa}, but there also are developments in other areas, such as electro-dynamical
applications \cite{Moon:2014gaa} and mathematical consistency \cite{Bischoff:2009re}. 
The four-dimensional CFT considered in this article begins with the question of
whether it is possible to develop a quaternion-parametrized four-dimensional conformal field theory (QCFT).
A quaternionic scalar field theory is already known \cite{Giardino:2012ti}, and so it would be seem a good idea 
to elaborate a four-dimensional model by preserving many of the properties of the two-dimensional theory and increasing 
the number of degrees of freedom. In fact, the quaternionic four-dimensional case has more constraints than the complex
two-dimensional case, meaning that several results that may be obtained in the two-dimensional CFT based on symmetry arguments
cannot be obtained in QCFT. Examples include correlation functions, creation and annihilation operators and the 
Hilbert space. However, the concept of generalizing two-dimensional CFT in terms of quaternion-parametrized four-dimensional 
CFT was not originally ours. \cite{Evans:1992az} reviews several attempts to achieve this aim for self-dual
Yang-Mills theories, and another example of this kind is provided by \cite{Popov:1998fb}. One serious restriction for constructing a 
quaternionic
four-dimensional theory is the severely limiting quaternion analyticity, which only admits affine quaternion functions in 
the left-derivative class \cite{Sudbery:1979qta}. A less restrictive quaternionic analyticity has been used for building 
a four-dimensional $\sigma-$model with applications in gauge theories \cite{Gursey:1979tu}.

In fact, the use of quaternions for generalizing complex-based theories is not straightforward, and quaternionic theories may not 
necessarily recover complex results by canceling the pure quaternionic variables. There are solutions for the Dirac equation
for the quaternionic step potential that have no complex limit \cite{DeLeo:2013xfa,DeLeo:2015hza}, and the 
quaternionic Dirac square well also has no counterpart within the complex limit \cite{Giardino:2015iia}. 
These questions enable us to suppose that using quaternions as a substitute for complex numbers may generate theories that are 
significantly different from  the complex parametrized theories that have inspired them. 

This article presents another attempt to test the feasibility of using quaternions to build a four-dimensional CFT. The
novelty here is that the conformal transformations are quaternion holomorphic functions (QHF). We believe that this proposal is the closest
QCFT  to a CCFT that has ever been built. The resulting quaternionic theory is more restrictive than the complex theory, first of all
because it has a more restrictive finite dimensional symmetry group, while the CCFT has an infinite dimensional Lie
algebra that parametrizes its symmetries \cite{Giardino:2015dza}. We decided to avoid a trivial theory by imposing quaternionic analyticity only 
on conformal symmetry transformations, and not on quaternionic functions that suffer the transformation. This choice permits
a wider class of quaternionic functions to be employed in the theory, and thus a QCFT has been formally built for holomorphic symmetry transformations
comprising the quantization of the fields and the operator product expansions.

The article is organized as follows, Section \ref{S2} presents the QHF's and the constraints they impose on conformal transformations and on
the stress tensor; in Section \ref{S3} the quantization of quaternionic primary fieds is discussed, and in Section \ref{S4} the 
operator product expansions (OPE's) are obtained for the stress tensor and for the quaternionic primary field. Section \ref{C} 
presents our conclusions and future perspectives for physical applications. Furthermore, Appendix \ref{A1} presents the quaternionic
parametrizations and metric tensors, and Appendix \ref{A2} comprises the symmetry algebra of M\"obius transformations for QHF's.

\section{Conformal Invariance \label{S2}}

A conformal transformation on a $(b+d)-$dimensional flat space $\mathbbm{R}^{b,d}$ is achieved through a coordinate change 
$x\to x'$  so that the metric tensor $\eta_{\mu\nu}$ transforms by a global scale
factor $H(x)$, so that 
\begin{equation}\label{cmt}
\eta_{\kappa\lambda}(x')\,\partial_\mu x^{\prime^\kappa}\,\partial_\nu x^{\prime\lambda}=H(x)\,\eta_{\mu\nu}(x),
\end{equation}
and the summation of indexes is implicit. In order to determine the
symmetry algebra of conformal transformations, we perform an infinitesimal transformation 
\begin{equation}
x'^\mu=x^\mu+\varepsilon^\mu(x),
\end{equation}
which applied on (\ref{cmt}) leads to 
\begin{equation}\label{ict}
\partial_\mu\varepsilon_\nu+\partial_\nu\varepsilon_\mu=\frac{2}{D} (\partial\cdot\varepsilon)\,\eta_{\mu\nu}\qquad\mbox{so that}
\qquad \partial\cdot\varepsilon=\eta^{\mu\nu}\partial_\mu\varepsilon_\nu.
\end{equation}
Higher order terms on $\epsilon_\mu$ were discarded, and $D=b+d$ is the dimension of the space.
A four dimensional Euclidean metric, where $\eta_{\mu\nu}=\delta_{\mu\nu}$ and $D=4$, implies that an 
Euclidean infinitesimal conformal transformation (\ref{ict}) changes to
\begin{equation}\label{eps_comp}
\partial_\mu\varepsilon_\nu=\frac{1}{4} \partial\cdot\varepsilon\qquad\mbox{for}\qquad \mu=\nu
\qquad\mbox{and}\qquad\partial_\mu\varepsilon_\nu=-\partial_\nu\varepsilon_\mu\qquad\mbox{for}\qquad \mu\neq\nu.
\end{equation}
Then we impose that a four dimensional conformal transformation
may be organized as a quaternion valued function $\mathcal{E}$ so that
\begin{equation}\label{epsilon_ex}
\mathcal{E}= \epsilon_0 +\epsilon_1i+\epsilon_2j +\epsilon_3k, 
\end{equation}
where $i,\,j$ and $k$ are the associative and anti-commutative complex units of quaternions, which obey
\[
i^2=j^2=k^2=-1\qquad\mbox{and}\qquad ijk=-1.
\] 
Using the symplectic notation, quaternions are written in complex components, so that
\begin{equation}\label{symp}
q=z+\zeta j \qquad\mbox{for}\qquad z=x_0+x_1i\qquad\mbox{and}\qquad \zeta=x_2+x_3i.
\end{equation}
Consequently, the quaternion evaluated function (\ref{epsilon_ex}) will be
\begin{equation}\label{epsilon_sym}
\mathcal{E}=\mathcal{E}_0+\mathcal{E}_1j,\qquad\mbox{so that}\qquad\mathcal{E}_0=\varepsilon_0+i\varepsilon_1\qquad\mbox{and}\qquad \mathcal{E}_1=\varepsilon_2+i\varepsilon_3.
\end{equation}
Now, we will assume a further constraint, that $\mathcal{E}$ is a holomorphic quaternion function. This class of 
function is very restrict, and admits only affine quaternion functions, so that 
\begin{equation}\label{Eqhf}
 \mathcal{E}=qa+b
\end{equation}
for $a,\;b$ and $q$ quaternionic. In symplectic notation (\ref{epsilon_sym}), $\mathcal{E}$ must obey some 
constraints in order to be holomorphic \cite{Sudbery:1979qta,Deavours:1973qtc}, namely
\begin{equation}\label{CR_pre}
\partial_0 \mathcal{E}_0=-i\partial_1 \mathcal{E}_0=\partial_2\bar{\mathcal{E}}_1=i\partial_3 \bar{\mathcal{E}_1},
\qquad\partial_0 \bar{\mathcal{E}_1}=i\partial_1 \bar{\mathcal{E}_1}=-\partial_2 \mathcal{E}_0=i\partial_3 \mathcal{E}_0
\end{equation}
and furthermore
\begin{equation}
\partial_z \mathcal{E}_0=\partial_{\bar\zeta}\bar{\mathcal{E}_1}\qquad\mbox{and}\qquad\partial_\zeta \mathcal{E}_0=-\partial_{\bar z}\bar{\mathcal{E}_1}.
\end{equation}
From (\ref{CR_pre}) and (\ref{epsilon_sym}), we see that the quaternionic function $\mathcal{E}$ will be
holomorphic if constrained by the previous condition (\ref{eps_comp}) and by the imposition of
\begin{equation}\label{constr}
\partial_0\varepsilon_1=-\partial_2\varepsilon_3,\qquad\partial_0\varepsilon_2=\partial_1\varepsilon_3,\qquad
\partial_0\varepsilon_3=-\partial_1\varepsilon_2\qquad\mbox{and}\qquad \partial_\mu\epsilon_\mu=\partial_\nu\epsilon_\nu.
\end{equation}
The situation is quite analogous to the two-dimensional case, where the Cauchy-Riemann conditions were generated
instead of (\ref{eps_comp}) and (\ref{constr}). There, holomorphic complex valued 
functions are the natural choice for two-dimensional conformal field theories. The quaternion case, on the other
hand, seems not to be most general case, because of the additional constraint (\ref{constr}). We might imagine that 
some class of quaternion function could generate (\ref{eps_comp}) either without further assumptions  
or by adopting less restrict conditions. At this moment, we cannot say whether this is possible or not, and the
quaternion holomorphic function appears to be the only feasible way to develop a quaternion parametrized conformal
field theory. By way of example, an holomorphic function $\mathcal{F}(q)=q+\mathcal{E}(q)$ applied on
a four dimensional metric leads to a conformal transformation
\begin{equation}
ds^2=dq\,d\bar q=\left|\frac{d\mathcal{F}}{dq}\right|^2 dq\,d\bar q,
\end{equation}
with $\left|\frac{d\mathcal{F}}{dq}\right|^2$ as scale factor and  where 
$d\bar q\to\overline{\frac{d\mathcal{F}}{dq}dq\,}=d\bar q\,\overline{\frac{d\mathcal{F}}{dq}}$ were used.

The symmetry algebra $\mathfrak{g}$ of the group $\mathsf{G}$ of transformations on left-derivative quaternion holomorphic
 functions like (\ref{Eqhf}) were studied at \cite{Giardino:2015dza}.
The $\mathsf{su}(2,\,\mathbb{C})$ algebra is generated by the $\{g_1,\,g_2,\,g_3\}$ sub-algebra of $\mathfrak{g}$. We
associate these operators with rotations on a two sphere, the operator $g_4$ generates dilations and $g_5$ and $g_6$ generate
translations. This algebra has an important difference when compared to the conformal algebra of two-dimensional theories, 
because the two dimensional case has an infinite dimensional algebra, the Witt algebra. In the two-dimensional case, the 
r\^ole of $\mathfrak{g}$ is played by the algebra of the  $\mathsf{SL}(2,\,\mathbb{C})/\mathbb{Z}_2$ group. The Witt algebra
is infinite dimensional and admits a central extension. On the other hand, $\mathfrak{g}$ is finite dimensional,
and consequently does not admit a central extension.

\subsection{Quaternion primary fields}

In two-dimensional complex CFT (CCFT), there exists a terminology for the physical fields according to their conformal properties. 
Complex holomorphic fields $\Phi(z)$ are called chiral and complex anti-holomorphic fields $\bar \Phi(\bar z)$ are called 
anti-chiral. For quaternion functions, we will not use the term holomorphic and anti-holomorphic, because these definitions
are too restrictive for quaternions, but we can loosely call $\Lambda(z,\,\zeta)$ chiral and $\bar\Lambda(\bar z,\,\bar\zeta)$
anti-chiral. We define a conformal transformation on a quaternionic field $\Lambda(q,\,\bar q)$ in terms of a transformation $\mathcal F$ on its coordinates 
as
\begin{equation}\label{prifield}
\Lambda(q,\,\bar q)=\overline{\left(\frac{\partial\mathcal{F}}{\partial q}\right)}^{\bar \ell}\left(\frac{\partial\mathcal{F}}{\partial q}\right)^\ell
\Lambda\left(\mathcal{F}(q),\,\overline{\mathcal{F}(q)}\right),
\end{equation}
so that the pair $(\ell,\,\bar\ell)$ is the conformal weight, which are supposed to be real. Complex conformal weights are found for particle
generating models \cite{Chatterjee:2015pha,Chen:2012zn,Pomoni:2008de}, and we will not consider here because of the non-commutativity of 
quaternions. Furthermore, we see that if $\ell\neq \bar \ell$, then there is an ambiguity in the ordering of the factors, and then we avoid this
problem imposing $\ell=\bar\ell$, and thus the product of the quaternion factors is real, precluding any ordering problem.
Note that $\Lambda$ needs not to be quaternion holomorphic. It may be a real
function parametrized by quaternions, like the quaternionic norm. On the other hand, we are dealing with quaternion holomorphic $\mathcal F$,
and then the conformal factors on (\ref{prifield}) are constant. 

Now we want to determine the change suffered on the field $\Lambda$ caused by
an infinitesimal quaternionic transformation 
\begin{equation}
\mathcal{F}=q+\mathcal E\qquad\mbox{so that}\qquad \mathcal E=q\,\epsilon+\delta,
\end{equation}
where $|\mathcal E|\ll 1$, $\epsilon=\epsilon_z+\epsilon_\zeta j$ and $\delta=\delta_z+\delta_\zeta j$.
We want to expand (\ref{prifield}) in a power series, but remembering that $\Lambda$ is not holomorphic, we cannot
obtain such a series using $q$ and $\bar q$ as variables. Then, we will use $z$, $\bar z$, $\zeta$ and $\bar \zeta$ as variables.
In symplectic coordinates, we get
\begin{equation}\label{Esymp}
\mathcal E=\mathcal E_z+\mathcal E_\zeta j= z\,\epsilon_z-\zeta\,\epsilon_{\bar\zeta}+\delta_z+\big(\zeta\epsilon_{\bar z}+z\epsilon_\zeta+\delta_\zeta\big)j,
\end{equation}
where $\epsilon_{\bar z}=\bar\epsilon_z$. In order to obtain the expansion, we write the operators
\begin{equation}
\hat{\mathcal E}=\big(z\,\epsilon_z-\zeta\,\epsilon_{\bar\zeta}+\delta_z\big)\partial_z+\big(\zeta\epsilon_{\bar z}+z\epsilon_\zeta+\delta_\zeta\big)\partial_\zeta,
\end{equation}
where $j\to\partial_\zeta$ and $1\to\partial_z$ have been used, and $\hat{\bar{\mathcal E}}$ may be obtained from the complex conjugate.
Thus, using (\ref{prifield}) and the approximated expansions
\begin{equation}
|\partial_q\mathcal F|^{2\ell}\approx 1+\ell\left(\epsilon_z+\epsilon_{\bar z}\right)\qquad\Lambda\approx \left(1+\hat{\mathcal E}+\hat{\bar{\mathcal E}}\right)\Lambda,
\end{equation}
we obtain
\begin{equation}\label{deltaL}
\delta \Lambda\approx\Big[\big(z\,\epsilon_z-\zeta\,\epsilon_{\bar\zeta}+\delta_z\big)\partial_z+
\big(\bar z\,\epsilon_{\bar z}-\bar\zeta\,\epsilon_{\zeta}+\delta_{\bar z}\big)\partial_{\bar z}+\ell(\epsilon_z+\epsilon_{\bar z})\Big]\Lambda.
\end{equation}
This result has important consequences, because there are no derivatives depending neither on $\zeta$ nor on $\bar\zeta$, and 
thus the symmetry operators depending on these derivatives does not contribute to the conserved charge. In order to build the 
conserved charge, we now discuss the stress tensor.
\subsection{The stress tensor}
The conserved charge associated to a symmetry is obtained from a conserved current $j_\mu$, so that 
\begin{equation}
j_\mu=T_{\mu\nu}\mathcal E^\nu,
\end{equation}
with $T_{\mu\nu}$ the symmetric stress tensor. For conformal theories, the conformal stress tensor
is always traceless. In four-dimensional symplectic coordinates, we have
\begin{equation}
 T_\mu^{\;\mu}=T_{z\bar z}+T_{\zeta\bar\zeta}=0,
\end{equation}
where the quaternion symplectic metric (\ref{smetric}) has been used. Another property of the stress tensor in symplectic 
coordinates is 
\[
\overline T_{\mu\nu}=T_{\bar\mu\bar\nu}
\]
which follows directly from $T_{\mu\nu}=\partial_\mu x^\alpha\partial_\nu x^\beta T_{\alpha\beta}$ with complex $\mu,\,\nu$ and real
$\alpha$, $x^\alpha$ and $T_{\alpha\beta}$. These conditions mean that from the ten  components of $T_{\mu\nu}$,
there are only five independent components, namely $T_{zz}$, $T_{\zeta\zeta}$, $T_{z\zeta}$, $T_{z\bar\zeta}$ and 
$T_{z\bar z}$. The $T_{z\bar z}$ component is identical to its complex conjugate, and consequently it is real. Then $T_{\mu\nu}$ has 
the nine degrees of freedom coming from the real stress tensor $T_{\alpha\beta}$, as expected. We may obtain the conserved charge in symplectic 
coordinates (\ref{sympcoords}) from the conserved current, so that
\begin{equation}
j_0=j_z+j_{\bar z}\qquad\mbox{where}\qquad  j_z=2\left(T_{z\bar z}\mathcal E_z+T_{z z}\mathcal E_{\bar z}+T_{z\bar\zeta}\mathcal E_\zeta
+T_{z\zeta}\mathcal E_{\bar\zeta}\right)
\qquad\mbox{and}\qquad j_{\bar z}=\bar j_z.
\end{equation}
Using (\ref{Esymp}) we obtain
\begin{equation}\label{jzjota}
j_z+j_{\bar z}=2\left[\left(z\epsilon_z-\zeta\epsilon_{\bar\zeta}+\delta_z\right)T_1+\left(z\epsilon_\zeta+\zeta\epsilon_{\bar z}+\delta_{\zeta}\right)T_2 + C.C.\right]
\qquad\mbox{where}\qquad T_1=T_{z\bar z}+T_{\bar z\bar z}\qquad\mbox{and}\qquad T_2=T_{z\bar\zeta}+T_{\bar z\bar\zeta}.
\end{equation}
The conserved current is obtained from the volume integral
\begin{equation}\label{cq}
Q=\int dv j_0=\int dv \left(j_z+j_{\bar z}\right).
\end{equation}
The symmetry transformation on $\Lambda$ is given by
\begin{equation}\label{delamb}
 \delta\Lambda=[Q,\,\Lambda]=\int dv\,\big[j_z+j_{\bar z},\,\Lambda\big],
\end{equation}
where the volume element $dv$ is real an may be factored out. 
In order to calculate the volume integral, we avail ourselves of the quaternion parametrization from appendix \ref{A1}, so that
\begin{equation}\label{polarq}
q=\cos\theta e^{\tau+i\phi}+\sin\theta e^{\tau+i\varphi}j,
\end{equation}
where $\theta\in[0,\,\pi/2]$ maintains the positivity of the radii of each complex component of (\ref{polarq}). 
This parametrization maps a four-dimensional cylinder to a quaternionic hyper-plane. We identify $\tau$ as a time coordinate, 
and we further consider $q=e^z+e^\zeta j$, so that
\begin{equation}
 z=\tau+\ln\cos\theta+i\phi,\qquad\zeta=\tau+\ln\sin\theta+i\varphi,
\end{equation}
and the radii of the complex parts in $q$ are not identical, namely $|e^z|\neq |e^\zeta|$. Thus, the volume element in the 
quaternionic plane for a constant time is calculated as
\begin{equation}\label{dv}
 dv=\cot\theta d\theta d\phi d\varphi.
\end{equation}
From (\ref{cq}) we observe the relationship between the conserved charge $Q$ and the conserved current $j_\mu$, and consequently
the correspondence between the stress tensor and the conserved charge. Now we want to determine the relation between the
conserved charge and the symmetry operators of the QHT. The symmetry operators $x_i$ that generate the QHT \cite{Giardino:2015dza}
are related to the infinitesimal parameters as
\begin{equation}\label{xcorr}
\delta_z\mapsto\partial_z=x_1,\qquad\epsilon_z\mapsto z\partial_z=x_2,\qquad\epsilon_{\bar\zeta}\mapsto \zeta\partial_z=x_3,\qquad
\delta_\zeta\mapsto\partial_\zeta=x_4,\qquad\epsilon_{\bar z}\mapsto \zeta\partial_\zeta=x_5,\qquad\epsilon_\zeta\mapsto z\partial_\zeta=x_6,
\end{equation}
and then we want to express $Q$ in terms of $x_i$ as
\begin{equation}
 Q=\sum_iQ_{\,i}=\sum_i\epsilon_i\,x_i.
\end{equation}
However, from (\ref{deltaL}) we see that $x_4,\,x_5$ and $x_6$ do not contribute for $\delta\Lambda$, and then we set
$T_2=0$ in  (\ref{jzjota}). In order to employ the current to determine the conserved charge, we  define
\begin{equation}
 \mathtt{u}=k+\kappa j, \qquad\mbox{so that}\qquad k=e^{\tau+i\phi}\qquad\mbox{and}\qquad \kappa=e^{\tau+i\varphi},
\end{equation}
and using (\ref{xcorr}) we write
\begin{equation}\label{TpT}
T_1=\frac{\tan\theta}{2}T\qquad\mbox{so that}\qquad
T=\frac{2}{\pi}\frac{x_1}{k\kappa}+\frac{x_2}{k^2\kappa}+\frac{x_3}{k\kappa^2}.
\end{equation}
$T$ depend on $k$ and $\kappa$, but not on their complex conjugates, and thus it may be considered analogue
of chiral fields of the CT. We eliminate the dependence on $\theta$ through the integration
\begin{equation}\label{jotas}
J+\bar J=\intop_0^{\pi/2}d\theta\cot\theta\left(j_z+j_{\bar z}\right)
=\left(k\epsilon_z-\kappa\epsilon_{\bar\zeta}+\delta_z\right)T + C.C.
\end{equation}
The $\phi$ and $\varphi$ integrals may be changed to $k$ and $\kappa$ integrals over the radius $|k|=|\kappa|=e^\tau$, and then
\begin{eqnarray}\label{Qcharge}
Q=\oint\frac{dk}{2\pi i}\oint \frac{d\kappa}{2\pi i}\,\big( J+\bar J\,\,\big)=\epsilon_z\, x_2+\epsilon_{\bar\zeta}\,x_3+
\delta_z\, x_1.
\end{eqnarray}
The $x_i$ generators may be obtained by inverting (\ref{TpT}), so that
\begin{equation}\label{xi}
x_1=\oint \frac{dk\,d\kappa}{(2\pi i)^2}\, T,\qquad 
x_2=\oint \frac{dk\,d\kappa}{(2\pi i)^2}\,k\,T\qquad
x_3=\oint \frac{dk\,d\kappa}{(2\pi i)^2}\,\kappa\,T.
\end{equation}
The results show a very constrained model, where the quaternionic coordinate $\zeta$ has no effect on the transformaton of the primary
field $\Lambda$. In fact, this is coherent with the transformation law (\ref{deltaL}), where the $\partial_\zeta$ does not generate
transformations.

\section{Quantized fields\label{S3}}

In this section, we outline several formal aspects of quantizing a QCFT. The discussion is
formal because every specific case to be quantized must be considered independently, and the discussion below only presents
general aspects that must be followed.

As in the two-dimensional case, the quaternion parametrization (\ref{polarq})
relates dilations to time translations, and $q\to 0$ is associated to the infinite
past, where $\tau\to-\infty$, or equivalently $q\to 0$. We note that the infinite past
is valid only for both the complex variables going to zero. They may independently go to zero for specific values of $\theta$, and
this is not associated to the infinite past. Using $e^\tau=\rho$ in (\ref{polarq}), we obtain the operators
\begin{equation}
\partial_z=\frac{e^{-i\phi}}{2}\left(\cos\theta\partial_\rho-\frac{\sin\theta}{\rho}\partial_\theta-\frac{i}{\rho\cos\theta}\partial_\phi\right)\qquad
\partial_\zeta=\frac{e^{-i\varphi}}{2}\left(\sin\theta\partial_\rho-\frac{\cos\theta}{\rho}\partial_\theta-\frac{i}{\rho\sin\theta}\partial_\varphi\right),
\end{equation}
which permit us to obtain
\begin{equation}\nonumber
\rho\partial_\rho=z\partial_z+\zeta\partial_\zeta+\bar z\partial_{\bar z}+\bar\zeta\partial_{\bar\zeta},\qquad
\partial_\theta=\sqrt{z\bar z\zeta\bar \zeta}\left(-\frac{1}{z}\partial_z+\frac{1}{\zeta}\partial_\zeta-\frac{1}{\bar z}\partial_{\bar z}+\frac{1}{\bar\zeta}\partial_{\bar\zeta}\right),
\qquad
\partial_\phi=i\left(z\partial_z-\bar z\partial_{\bar z}\right),\qquad
\partial_\varphi=i\left(\zeta\partial_\zeta-\bar \zeta\partial_{\bar \zeta}\right).
\end{equation}
In terms of the generators (\ref{ggenerators}) of the algebra of $\mathsf G$, we can write
\begin{equation}
\rho\partial_\rho=-\left(g_4+\bar g_4\right)\qquad \partial_\phi=i\left(g_3-g_4-\bar g_3+\bar g_4\right)
\qquad \partial_\varphi=i\left(\bar g_3+\bar g_4-g_3-g_4\right)
\end{equation}
At this point, we observe a difference between the two-dimensional CCFT and the four-dimensional QCFT.  
In two dimensions the time operator is associated to the real dilation and the space translations may be written in terms of a 
linear combination of symmetry operators. In four dimensions, this is no longer observed because  the $\theta$ coordinate cannot
be written in terms of the symmetry operators of $\mathfrak{g}$. Furthermore, the direct sum of the symmetry algebra $\mathfrak{g}$ and
the algebra generated by the complex conjugates of its operators, $\mathfrak{g}\oplus\bar{\mathfrak g}$, is not an algebra. In two-dimensions,
this direct sum is an algebra, and this very fact presents a big difference between the two cases, indicating that the four-dimensinal
case is more restrictive because it cannot avail itself of the full symmetry algebra. We note that these differences raise due to the
complex conjugacy of quaternions, were $\overline{(z,\,\zeta)}\to(\bar z,\,-\zeta)$.

The second aspect to be considered is defining a primary field to be quantized, and thereto we propose the expansion
\begin{equation}\label{qlaur2}
\Lambda(z,\bar z,\,\zeta,\,\bar\zeta)=(q\bar q)^{-\ell}\sum_{\bar m,\,n,\,r,\,\bar s\in\mathbb{Z}}z^{-n}\,\bar z^{\,-\bar m}\,
\zeta^{-r}\,\bar \zeta^{\,-\bar s}\Lambda_{\bar m,\,n,\, r,\,\bar s},
\end{equation}
where $\Lambda_{\bar m,\,n,\, r,\,\bar s}$ are quaternionic constants, and quantization is obtained promoting the primary field (\ref{qlaur2})
to an operator $\hat\Lambda$. We define asymptotic past states using $\tau\to-\infty$ in (\ref{polarq}), so that $q\to 0$ and then
asymptotic past states may be written
\begin{equation}
 |\Lambda\rangle_\infty=\lim_{q\to 0}\hat\Lambda|0\rangle.
\end{equation}
Considering that all coordinates have the real factor $e^\tau$ which 
governs the approaching to zero, we eliminate negative powers on the variables of $\hat\Lambda$ by imposing
\begin{equation}
\hat\Lambda_{\bar m,n,\, r,\,\bar s}|0\rangle=0\qquad\mbox{for}\qquad \sigma+2\ell>0,\qquad\mbox{so that}\qquad\sigma=n+\bar m+r+\bar s
\end{equation}
The asymptotic state is then generated by the independent term of $\hat\Lambda$, so that $\sigma=0$.

A third important formal aspect of the quantized primary field is the Hermitian conjugate $\Lambda^\dagger$ of the primary field. Considering 
that Hermitian conjugacy changes the sign of the time coordinate, then $\tau\to-\tau$ in (\ref{polarq}). Remembering that 
$e^\tau=\sqrt{q\bar q}$, then
\begin{equation}
z^\dagger=\frac{z}{q\bar q},\qquad\zeta^\dagger=\frac{\zeta}{q\bar q}\qquad\mbox{and consequently}\qquad q^\dagger=\frac{q}{q\bar q}.
\end{equation}
We then see that Hermitian conjugacy is not a quaternion conformal transformation as defined in \cite{Giardino:2015dza}
because it implies in the inversion of the quaternionic coordinate.  We thus define
\begin{equation}
\Lambda^\dagger(z,\bar z,\,\zeta,\,\bar\zeta)=(q\bar q)^{\ell+\sigma}\sum_{\bar m,\,n,\,r,\,\bar s\in\mathbb{Z}}
\Lambda^\dagger_{\bar m,\,n,\, r,\,\bar s}\,z^{-n}\bar z^{\,-\bar m}\,\zeta^{-r}\,\bar \zeta^{\,-\bar s},
\end{equation}
such that the negative powers on the variables of $\hat\Lambda^\dagger$ are eliminated by imposing
\[
\hat\Lambda^\dagger_{\bar m,n,\, r,\,\bar s}|0\rangle=0\qquad\mbox{for}\qquad \sigma+2\ell<0,
\]
and then we have well-defined primary fields for quantizing.

\section{Operator products\label{S4}}
In two-dimensional CFT, the time coordinate is parametrized by the radial direction, so that $z=e^{x_0+ix_1}$. Thus, the time
ordering of an operator product like $\hat X(z)\hat Y(w)$ is determined by the relation between $|z|$ and $|w|$. In the quaternion
parametrization (\ref{polarq}) this is also true because $|q|=e^\tau$ and $\tau$ is identified with a time coordinate.
However, as $q=z+\zeta j$, the time direction cannot be identified  neither with of $|z|$ nor with $|\zeta|$. This is an important detail
in order to calculate the conserved charge and consequently the symmetry transformations. Using (\ref{Qcharge}), we rewrite 
(\ref{delamb}) as
\begin{equation}\label{deltaL2}
\delta\Lambda=\oint\frac{ dk}{2\pi i}\oint\frac{d\kappa}{2\pi i}\,\big[J+\bar J,\,\Lambda\big].
\end{equation}
We want to get some physical insight from the equality between (\ref{deltaL}) and (\ref{delamb}), namely
\begin{equation}\label{OPEdelta}
\int dv[j_z+j_{\bar z},\,\Lambda]
=\left[\epsilon_z\left(\ell+z\partial_z\right)-\epsilon_{\bar\zeta}\zeta\partial_z+\delta_z\partial_z+C.C.\right]\Lambda.
\end{equation}
The $\theta-$dependence may be eliminated from the right hand side of (\ref{OPEdelta}) using (\ref{jotas}), and the integral on
$\phi$ and $\varphi$ may be turned into $k$ and $\kappa$ integrals. However, the right hand side depends on $z$ and $\zeta$
variables, which explicitly depend on $\theta$. Our interpretation about this fact is that the $\theta-$coordinate does not have
significance on the conformal symmetry, and then we may discuss the conformal symmetry only in terms of the $\theta-$free coordinates
$k$ and $\kappa$.  Using (\ref{jotas}), we turn the commutator (\ref{OPEdelta}) into a time ordered product 
$\Theta$, so that
\begin{equation}\label{Tetas}
[J+\bar J,\,\Lambda]\mapsto\Theta\left[\big(J+\bar J\,\big)\Lambda\right]
=\left(k\epsilon_z-\kappa\epsilon_{\bar\zeta}+\delta_z\right)\Theta\big[T\Lambda\big]+
\left(\bar k\epsilon_{\bar z}-\bar\kappa\epsilon_\zeta+\delta_{\bar z}\right)\Theta\big[\overline T\Lambda\big]
\end{equation}
We may determine the operator product expansion by substitution (\ref{Tetas}) back in (\ref{OPEdelta}) and using the identities
\begin{equation}\label{ident}
 f(w)=\frac{1}{2\pi i}\oint dk\frac{f(k)}{k-w}\qquad\mbox{and}\qquad\partial_wf(w)=\frac{1}{2\pi i}\oint dk\frac{f(k)}{(k-w)^2},
\end{equation}
where $f(k)$ is a complex holomorphic function. Thus, we obtain
\begin{equation}\label{EPOlambda}
\Theta\big[T\,\Lambda\big]=\frac{1}{(k-w)(\kappa-\omega)}\partial_w \Lambda
+\frac{ \ell}{(k-w)^2(\kappa-\omega)}\Lambda.
\end{equation}
Using (\ref{xi}) and the $\mathfrak{x}$ algebra (\ref{xalgebra}), we calculate the OPE for $T$,
\begin{equation}\label{epoT}
\Theta\big[T(k,\,\kappa)T(w,\,\omega)\big]=2\frac{T(w,\,\omega)}{(k-w)^2\kappa}
-\frac{ T(w,\,\omega)}{(k-w)^2(\kappa-\omega)}.
\end{equation}
If we make $\Lambda=T$ in (\ref{EPOlambda}) we see that $T$ is not a primary field. Indeed, using (\ref{epoT}) and $\Lambda=T$ in
 (\ref{Tetas}) and (\ref{deltaL2})  $\Lambda=T$, we get
\begin{equation}
 \delta T= \epsilon_zT.
\end{equation}
Finally, we use the OPE (\ref{EPOlambda}) 
and (\ref{xi})  to calculate the commutators among the symmetry operators $x_i$ and
the primary field $\Lambda$,
\begin{equation}
[x_1,\,\Lambda]=\Lambda,\qquad [x_2,\,\Lambda]=\ell\Lambda-w\partial_w\Lambda,\qquad [x_3,\,\Lambda]=\omega\partial_w\Lambda.
\end{equation}
In CCFT the commutation relations between primary fields and the symmetry operators are useful for 
defining creation and annihilation operators. In the QCFT discussed here this must still be the case, but as we were unable
to find a general expression for them, we leave this task as a direction of research to be studied in specific examples in
the future. Another quantity that is calculated in general in CCFT are correlation functions, but these computations may not be 
trivial in four dimensional \cite{Vos:2014pqa,Elkhidir:2014woa}, and then we postpone this issue for a future study.

\section{Conclusion\label{C}}

In this article we presented a four-dimensional CFT parametrized with quaternions (QCFT) and whose conformal transformations
are restricted to quaternion holomorphic functions (QHF). We have shown that such a theory may be quantized and that operator
product expansions (OPE's) may also be defined. This theory must of course now be tested in order to see its properties
in building concrete examples of physical theories, and the well known examples for CCFT are the most important candidates.
The study of scalar and fermionic fields are certainly interesting future directions for research, but even applications to string 
theory, the Thirring model and super-symmetric theories may be tried. Only after these applications is that the utility of 
QCFT may be correctly evaluated. At this point, the construction presented in this article is only the outline of a mathematical 
theory.

\section*{ACKNOWLEDGEMENTS}
Sergio Giardino receives the financial grant number 206383/2014-2 from the CNPq for his research and is grateful for the hospitality of Professor Paulo 
Vargas Moniz and the Centre for Mathematics and Applications of the Beira Interior University. 

\appendix

\section{Metric tensor \label{A1}}
A four-dimensional Euclidean metric $g_{\mu\nu}$ has different expressions using different quaternion variables. For
quaternion extended variables, 
\begin{equation}
 q=x_0+x_1i+x_2j+x_3k\qquad\Rightarrow\qquad g_{\mu\nu}=\delta_{\mu\nu}.
\end{equation}
Different coordinate systems lead to changes in the metric tensor. Using the Euclidean coordinates in terms of the symplectic ones 
(\ref{symp}), we get
\begin{equation}\label{sympcoords}
x_0=\frac{q+\bar q}{2}=\frac{z+\bar z}{2},\qquad x_1=\frac{\bar q i+i q}{2}=\frac{z-\bar z}{2i},\qquad 
x_2=\frac{\bar q j-j q}{2}=\frac{\zeta+\bar\zeta}{2},\qquad x_3=\frac{\bar q k-kq}{2}=\frac{\zeta-\bar\zeta}{2i}.
\end{equation}
The metric tensor shall be  calculated using
\begin{equation}\label{genmetric}
g_{\mu\nu}=\partial_\mu x^\kappa\partial_\nu x^\lambda G_{\kappa\lambda}.
\end{equation}
Using the Euclidean metric, so that $G_{\mu\nu}=\delta_{\mu\nu}$, and defining $g_{\mu\nu}=h_{\mu\nu}$, we obtain
\begin{equation}\label{smetric}
h_{\mu\nu}=\partial_\mu x^\kappa\partial_\nu x^\lambda\delta_{\kappa\lambda}\qquad\mbox{so that}\qquad
h_{\mu\nu}=\frac{1}{2}\left[
\begin{array}{llll}
0\;\;1\;\;0\;\;0\\
1\;\;0\;\;0\;\;0\\
0\;\;0\;\;0\;\;1\\
0\;\;0\;\;1\;\;0
\end{array}
\right].
\end{equation}
We call $h_{\mu\nu}$ the quaternion symplectic metric (QSM). Let us see some other ways of obtaining it using different coordinate
systems. In polar coordinates,
\begin{equation}\label{qpolar}
 q=\rho\left(\cos\theta e^{i\phi}+\sin\theta e^{i\varphi}j\right)\qquad\Rightarrow\qquad g_{\mu\nu}=
\left[
\begin{array}{llll}
1\;\;\;0\;\;\;\;0\qquad\qquad 0\\
0\;\;\;\rho^2\;\,0\qquad\qquad 0\\
0\;\;\;0\;\;\;\;\rho^2\cos^2\theta\;\;\;\, 0\\
0\;\;\;0\;\;\;\;0\qquad\qquad\rho^2\sin^2\theta
\end{array}
\right]=
\left[
\begin{array}{llll}
g_{\rho\rho}\;\;\;\;0\qquad0\qquad 0\\
0\qquad g_{\theta\theta}\;\;\;\;0\qquad 0\\
0\qquad 0\qquad g_{\phi\phi}\;\;\;\; 0\\
0\qquad 0\qquad0\qquad g_{\varphi\varphi}
\end{array}
\right],
\end{equation}
and
\begin{equation}\label{polar}
 \rho=\sqrt{x_0^2+x_1^2+x_2^2+x_3^2},\qquad\cos\theta=\frac{\sqrt{x_0^2+x_1^2}}{\rho},\qquad
\cos\phi=\frac{x_0}{\sqrt{x_0^2+x_1^2}},\qquad\cos\varphi=\frac{x_2}{\sqrt{x_2^2+x_3^2}},
\end{equation}
Rewriting (\ref{spolar}) in symplectic coordinates,
\begin{equation}\label{spolar}
\rho=\sqrt{z\bar z+\zeta\bar\zeta},\qquad \phi=\frac{1}{2i}\ln\frac{z}{\bar z},\qquad \phi=\frac{1}{2i}\ln\frac{\zeta}{\bar \zeta}
\qquad\theta=\frac{1}{2}\arccos\frac{z\bar z-\zeta\bar\zeta}{z\bar z+\zeta\bar\zeta},
\end{equation}
we obtain the quaternion symplectic metric $h_{\mu\nu}$ using (\ref{genmetric}), (\ref{qpolar}) and (\ref{spolar}). 
The important case for conformal field theory is a variant of (\ref{qpolar}), where $\rho=e^\tau$
\begin{equation}\label{qpolar2}
 q=\cos\theta e^{\tau+i\phi}+\sin\theta e^{\tau+i\varphi}j\qquad\Rightarrow\qquad g_{\mu\nu}=
e^{2\tau}\left[
\begin{array}{llll}
1\;\;\;0\;\;\;0\;\qquad 0\\
0\;\;\;1\;\;\;0\;\qquad 0\\
0\;\;\;0\;\;\;\cos^2\theta\; 0\\
0\;\;\;0\;\;\;0\;\qquad\sin^2\theta
\end{array}
\right]=
\left[
\begin{array}{llll}
g_{\tau\tau}\;\;\;\;0\qquad0\qquad 0\\
0\qquad g_{\theta\theta}\;\;\;\;0\qquad 0\\
0\qquad 0\qquad g_{\phi\phi}\;\;\;\; 0\\
0\qquad 0\qquad0\qquad g_{\varphi\varphi}
\end{array}
\right].
\end{equation}
where $z=\cos\theta e^{\tau+i\phi}$ and $\zeta=\sin\theta e^{\tau+i\varphi}$. We note the conformal symmetry of this metric, where
$e^{2\tau}$ is the conformal factor. On the other hand, from (\ref{qpolar2}) and using symplectic coordinates, we get
\begin{equation}
\tau=\frac{1}{2}\ln\left(z\bar z +\zeta\bar\zeta\right),\qquad \phi=\frac{1}{2i}\ln\frac{z}{\bar z},
\qquad \phi=\frac{1}{2i}\ln\frac{z}{\bar z},\qquad\theta=\arctan\sqrt{\frac{\zeta\bar\zeta}{z\bar z}}.
\end{equation}
Using this coordinate system, we obtain the metric given by (\ref{smetric}). It is important to note that always when a system
of coordinates is changed to the symplectic coordinates, the QSM will be obtained. However, we see that (\ref{qpolar2})
has conformal properties, and because of this it must be choosen for QCFT. As a last example, using
\begin{equation}\label{qpolar3}
q=e^{\chi+i\phi}+e^{\xi+i\varphi}j \qquad\mbox{where}\qquad r=z\bar z\qquad\mbox{and}\qquad\rho=\zeta\bar\zeta.
\qquad\Rightarrow\qquad g_{\mu\nu}=
\left[
\begin{array}{llll}
e^{2\chi}\;\,0\;\;\;\;\;0\;\;\;\; 0\\
0\;\;\;\;\;e^{2\chi}\;\,0\;\;\;\; 0\\
0\;\;\;\;\;0\;\;\;\;\;e^{2\xi}\;\, 0\\
0\;\;\;\;\;0\;\;\;\;\;0\;\;\;\;e^{2\xi}
\end{array}
\right],
\end{equation}
we again obtain the QSM in symplectic coordinates, but  in (\ref{qpolar3}) the conformal symmetry is not evident, and then 
it will not be used to study the QCFT.

\section{Quaternion holomorphic transformation representations\label{A2}}

The quaternion holomorphic transformation \cite{Giardino:2015dza} has two important representions. The first is $\mathfrak{x}$, given by
\begin{equation}
 x_1=\partial_z,\qquad x_2=z\partial_z\qquad x_3=\zeta\partial_z \qquad x_4=\partial_\zeta,  \qquad x_5=\zeta\partial_\zeta,
\qquad x_6 =z\partial_\zeta.
\end{equation}
The Lie algebra for $\mathfrak{x}$ is written through the commutation relations
\begin{eqnarray}\label{xalgebra}
&&\nonumber [x_1,\,x_2]=x_1\\
&&\nonumber [x_1,\,x_3]=0\qquad [x_2,\,x_3]=-x_3\\
&&\nonumber [x_1,\,x_4]=0\qquad [x_2,\,x_4]=0 \qquad [x_3,\,x_4]=-x_1\\
&&\nonumber [x_1,\,x_5]=0\qquad [x_2,\,x_5]=0\qquad [x_3,\,x_5]=-x_3,\qquad [x_4,\,x_5]=x_4\\
&& [x_1,\,x_6]=x_4\;\;\;\;\;\;\; [x_2,\,x_6]=x_6\;\;\;\;\;\;\;\, [x_3,\,x_6]=x_5-x_2\;\;\;\;\;\; [x_4,\,x_6]=0\qquad [x_5,\,x_6]=-x_6
\end{eqnarray}
The second representaion used in this article is $\mathfrak{g}$, whose generators
\begin{eqnarray}\label{ggenerators}
g_1=x_3+x_6,\qquad g_2=i\left(x_6-x_3\right),\qquad g_3=x_2-x_5,\qquad g_4=-\left(x_2+x_5\right),\qquad g_5=x_1,\qquad g_6=-x_4
\end{eqnarray}
obey the algebra
\begin{eqnarray}\label{galgebra}
&&\nonumber [g_1,\,g_2]=2ig_3\\
&&\nonumber [g_1,\,g_3]=2ig_2\;\;\;\;\;[g_2,\,g_3]=2ig_1\\
&&\nonumber [g_1,\,g_4]=0\qquad\,\; [g_2,\,g_4]=0 \qquad\;\;\;\, [g_3,\,g_4]=0\\
&&\nonumber [g_1,\,g_5]=g_6\qquad [g_2,\,g_5]=ig_6\qquad [g_3,\,g_5]=-g_5,\;\;\;\,[g_4,\,g_5]=g_5\\
&& [g_1,\,g_6]=g_5\qquad [g_2,\,g_6]=-ig_5\;\;\;\;\; [g_3,\,g_6]=g_6\qquad\, [g_4,\,g_6]=g_6\qquad [g_5,\,g_6]=0
\end{eqnarray}
This representation is interesting because we see the $\mathsf{su}(2)$ algebra generated by $g_1,\,g_2$ and $g_3$ and the complex
dilation generated by $g_4$. $g_5$ and $g_6$ are associated with complex translations.

%
%
%
%

\bibliographystyle{unsrt} 
\bibliography{bib_qcft}

\end{document}